\documentclass[aps,epsfig,rotate,preprint]{revtex4}
\usepackage{amssymb}
\usepackage{latexsym}
\usepackage{amsfonts}
\usepackage{graphicx}
\usepackage{epsfig}
\usepackage{amsmath}
\usepackage{float}  
\usepackage{verbatim}

\begin{document}

\title{Indication of multiscaling in the volatility return intervals
of stock markets}

\author{Fengzhong Wang$^1$, Kazuko Yamasaki$^{1,2}$, Shlomo
Havlin$^{1,3}$ and H. Eugene Stanley$^1$}

\affiliation{$^1$Center for Polymer Studies and Department of Physics,
Boston University, Boston, MA 02215 USA\\$^2$Department of
Environmental Sciences, Tokyo University of Information Sciences,
Chiba 265-8501,Japan\\$^3$Minerva Center and Department of Physics,
Bar-Ilan University, Ramat-Gan 52900, Israel}

\date{30 July 2007 version ~~~ wyhs.tex}

\begin{abstract}

The distribution of the return intervals $\tau$ between price
volatilities above a threshold height $q$ for financial records has
been approximated by a scaling behavior. To explore how accurate is
the scaling and therefore understand the underlined non-linear
mechanism, we investigate intraday datasets of 500 stocks which
consist of the Standard \& Poor's 500 index. We show that the
cumulative distribution of return intervals has systematic deviations
from scaling. We support this finding by studying the $m$-th moment
$\mu_m \equiv \langle(\tau/\langle\tau\rangle)^m\rangle^{1/m}$, which
show a certain trend with the mean interval $\langle\tau\rangle$. We
generate surrogate records using the Schreiber method, and find that
their cumulative distributions almost collapse to a single curve and
moments are almost constant for most range of
$\langle\tau\rangle$. Those substantial differences suggest that
non-linear correlations in the original volatility sequence account
for the deviations from a single scaling law. We also find that the
original and surrogate records exhibit slight tendencies for short and
long $\langle\tau\rangle$, due to the discreteness and finite size
effects of the records respectively. To avoid as possible those
effects for testing the multiscaling behavior, we investigate the
moments in the range $10<\langle\tau\rangle\leq100$, and find the
exponent $\alpha$ from the power law fitting
$\mu_m\sim\langle\tau\rangle^\alpha$ has a narrow distribution around
$\alpha\neq0$ which depend on $m$ for the 500 stocks. The distribution
of $\alpha$ for the surrogate records are very narrow and centered
around $\alpha=0$. This suggests that the return interval distribution
exhibit multiscaling behavior due to the non-linear correlations in
the original volatility.
\end{abstract}


\maketitle

\section{Introduction}

The price dynamics of financial markets has long been a focus of
economics and econophysics research \cite{Mandelbrot63,Mantegna95,
Kondor99,Mantegna00,Takayasu97,Liu99,Weber07,Bouchaud03}. Studying the
volatility time series is not only crucial for revealing the
underlined mechanism of financial markets dynamics, but also useful
for traders. For example, it helps traders to estimate the risk and
optimize the portfolio~\cite{Bouchaud03, Johnson03}. The volatility
series is known to be long-term power-law correlated
\cite{Ding83,Wood85,Harris86,Admati88,Schwert89,Dacorogna93,Granger96,Pagan96,Liu97,Cizeau97,Cont98,Pasquini99}. To
better understand these correlations and characterize temporal scaling
features in volatilities, recently Yamasaki et al. \cite{Yamasaki05}
and Wang et al. \cite{Wang06,Wang07} studied the statistics of return
intervals $\tau$ between volatilities that are above a given threshold
$q$, which is an alternative way to analyze long-term correlated time
series (see Ref~\cite{Altmann05} and references therein). They find
that scaling and memory in the return intervals of daily and intraday
financial records are similar to that found in the climate and
earthquake data \cite{Bunde04,Bunde05,Livina05}.

Studies of financial records show that the scaling in the return
intervals distribution can be well approximated by a scaling function
\cite{Yamasaki05,Wang06,Wang07}. However, financial time series are
known to show complex behavior and are not of uniscaling
nature~\cite{Matteo07} and non-linear features~\cite{Cao92}. Recent
studies \cite{Ivanov04,Eisler06A, Eisler06B} of stock markets show
that the distribution of activity measure such as the intertrade time
has multiscaling behavior. Thus, a detailed analysis of the scaling
properties of the volatility return intervals is of interest. It might
improve our understanding of the return intervals statistics and shed
light on the underlined complex mechanism of the volatility. Our
analysis suggests that for all Standard \& Poor's index constituents,
the cumulative distributions of the return intervals depart slightly
but systematically from a single scaling law. We also find that the
moments $\mu_m\equiv\langle(\tau/\langle\tau\rangle)^m\rangle^{1/m}$
are consistent with the deviations from scaling. However, using the
corresponding surrogate records \cite{Schreiber96,Makse96,Schreiber00}
which remove the non-linearities, $\mu_m$ almost does not depend on
$\langle\tau\rangle$ and no deviation from scaling occur. Therefore,
our results suggest that non-linear correlations in the volatility
account for the deviations from a scaling law.

The paper is organized as follows: In section II we introduce the
database and define the volatility. In section III we discuss the
scaling and investigate the deviations from scaling in the cumulative
distributions of the return intervals. We also describe the stretched
exponential form suggested for the distribution and the generation of
the surrogate records. Section IV deals with the moments of the return
intervals. We quantify the deviation from the scaling that exhibits
multiscaling behavior. We simulate the return intervals with different
sizes and show the finite size effect for long
$\langle\tau\rangle$. We also study the discreteness effect for short
$\langle\tau\rangle$ and explore the relation between the moment and
its order. In Section V we present a discussion.

\section{Database}

We analyze the Trades And Quotes (TAQ) database from New York Stock
Exchange (NYSE), which records every trade for all the securities in
United States stock markets. The duration is from Jan 1, 2001 to Dec
31, 2002, which has a total of 500 trading days. We study all 500
companies which consist of the Standard \& Poor's 500 index (S\&P 500)
\cite{Note1}, the benchmark for American stock markets. The volatility
is defined the same as in \cite{Wang06}. First we take the absolute
value for the logarithmic price change, then remove the intraday
U-shape pattern, and finally normalize it with the standard
deviation. Here the price is the closest tick to a minute mark. Thus
the sampling time is 1 minute and a trading day usually has 391 points
after removing the market closing hours. For each stock, the size of
dataset is about 200,000 records.

\section{Scaling in Return Intervals}

The probability density function (PDF) for the return intervals $\tau$
of the financial volatilities is well-approximated by the following form,
\begin{equation}
P_q(\tau)=\frac{1}{\langle\tau\rangle}f(\tau/\langle\tau\rangle),
\label{pdf.eq}
\end{equation}
as analyzed by Yamasaki et al.~\cite{Yamasaki05} and Wang et
al. \cite{Wang06,Wang07}. Here $\langle\cdot\rangle$ stands for the
average over the dataset and $\tau/\langle\tau\rangle$ depends on the
threshold $q$. It was suggested that the scaling function can be
approximated by a stretched exponential
\cite{Yamasaki05,Wang06,Wang07},
\begin{equation}
f(x)=c e^{-(ax)^\gamma}
\label{scaling.eq}
\end{equation} for financial records, which is consistent with other
long-term correlated records
\cite{Altmann05,Bunde04,Bunde05,Livina05}. Here $a$ and $c$ are
fitting parameters and $\gamma$ is the exponent characterizing the
long-term correlation \cite{Altmann05,Bunde04,Bunde05,Livina05}. From
the normalization of PDF follows \cite{Note2},
\begin{equation}
1 = \int^\infty_0 P_q(\tau) d\tau = \int^\infty_0
\frac{1}{\langle\tau\rangle} c e^{-(a\tau/\langle\tau\rangle)^\gamma}
d\tau.
\label{normal.eq}
\end{equation}
From the definition of $\langle\tau\rangle$ follows,
\begin{equation}
\langle\tau\rangle = \int^\infty_0 \tau\cdot P_q(\tau)
d\tau=\int^\infty_0 \tau\cdot \frac{1}{\langle\tau\rangle} c
e^{-(a\tau/\langle\tau\rangle)^\gamma}d\tau.
\label{mean.eq}
\end{equation}
Thus, using Eqs. (\ref{normal.eq}) and (\ref{mean.eq}), the parameters
$a$ and $c$ can be expressed by $\gamma$,
\begin{eqnarray}
a & = & \Gamma(2/\gamma)/\Gamma(1/\gamma), \nonumber\\
c & = & \gamma a/\Gamma(1/\gamma) = \gamma \Gamma(2/\gamma)/\Gamma(1/\gamma)^2.
\label{parameter.eq}
\end{eqnarray}
Here $\Gamma(a)\equiv\int^\infty_0 t^{a-1}e^{-t}dt$ is the Gamma
function. Thus, if the stretched exponential distribution is valid for
the scaled interval $\tau/\langle\tau\rangle$, it is completely
determined by $\gamma$. For $\gamma=1$, the record has no long-term
correlations and the return interval distribution indeed follows an
exponential distribution, represented by a Poissonian statistics, as
expected.

 Though the scaling in the return intervals distribution is a good
approximation, we find slight deviations that as shown below are
attributed to non-linear features. To explicitly explore the quality
of the scaling in return interval distributions, we study all S\&P 500
constituents and show the results of four representative stocks,
Citigroup (C), General Electric (GE), Coca Cola (KO) and Exxon Mobil
(XOM). All other stocks studied here usually show similar
features. First, we examine the cumulative distribution of the scaled
intervals.
\begin{equation}
D(\tau/\langle\tau\rangle)\equiv\int^\infty_{\tau}
P_q(\tau)d\tau=\int^\infty_{\tau/\langle\tau\rangle} f(x)dx.
\end{equation}

\begin{figure*}
\begin{center}
   \includegraphics[width=0.65\textwidth, angle = -90]{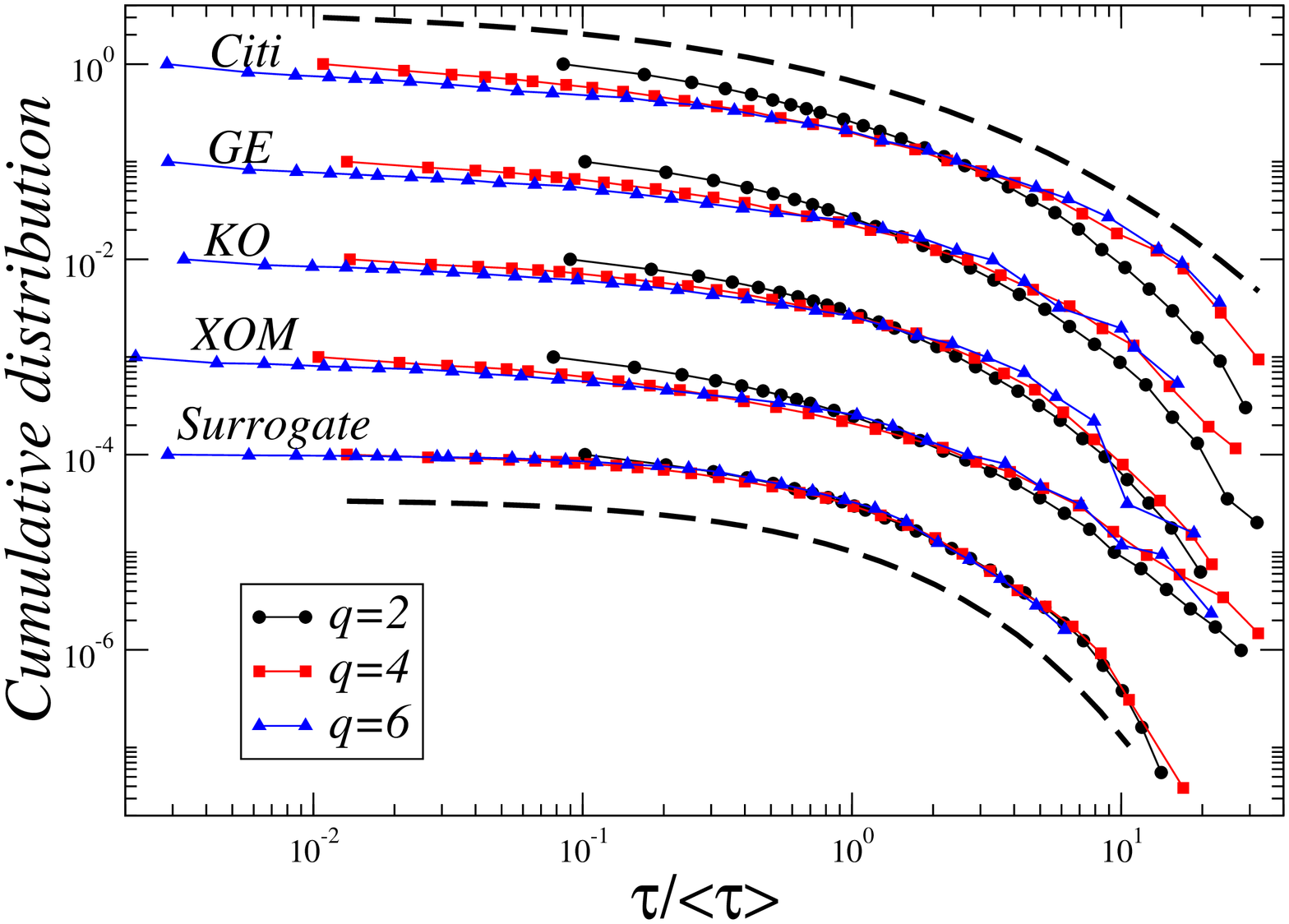}
\end{center}
\caption{(Color online) Cumulative distribution of the scaled
intervals $\tau/\langle\tau\rangle$ for stock C, GE, KO, XOM and GE's
surrogate. Symbols are for three thresholds $q=2$ (circles), $4$
(squares) and $6$ (triangles) respectively. As two examples, we fit
the cumulative distribution of stock C ($q=2$) and the surrogate
($q=2$) to a stretched exponential distribution
(Eq.~(\ref{scaling.eq})) with exponent $\gamma=0.25$ and $\gamma=0.50$
correspondingly. Except that for stock C, all symbols and curves are
vertically shifted for better visibility. For the surrogate records,
symbols collapse almost perfectly to one curve. However, all original
data exhibit similar systematic deviations from the scaling. This
suggests the non-linear correlations in the volatility series affects
the scaling in its return intervals.}
\label{Fig1}
\end{figure*}

If the scaling function $f(\tau/\langle\tau\rangle)$ is valid, the
cumulative distributions should also collapse to a single curve.
Otherwise, the cumulative distributions, which integrate deviations,
may show clearer deviations from scaling. Indeed, in Fig.~\ref{Fig1}
we show cumulative distributions for three thresholds $q=2$, $4$ and
$6$. Note that the volatility is normalized by its standard deviation,
the threshold $q$ is in units of standard deviations and therefore
$q=6$ is a quite large volatility. It is clearly seen that those
distributions are close to each other but do not collapse to a single
curve. More important, they show apparent deviations from the scaling,
which are systematic with the threshold. For small scaled intervals
($\tau/\langle\tau\rangle<1$), the cumulative distribution decreases
with $q$, while for large scaled intervals
($\tau/\langle\tau\rangle>1$), it increases with $q$ \cite{Note3}. In
other words, the scaled interval prefers to be larger for higher
threshold. This systematic trend suggests multiscaling in the return
intervals, which might be related to the non-linear correlations in
the volatility.

To better understand the systematic trends and test if it is not due
to finite size effect or discreteness of minutes, we also measure the
cumulative distribution of return intervals for surrogate records of
volatilities using the Schreiber method
\cite{Schreiber96,Makse96,Schreiber00} where non-linearities are
removed. For a given time series, we store the power spectrum and
randomly shuffle the sequence, then we apply the following
iterations. Each iteration consists of two consecutive steps:

(i) We perform the Fourier transform of the shuffled series, replace
its power spectrum with the original one, then take the inverse
Fourier transform to achieve a series. This step enforces the desired
power spectrum to the series, while the distribution of volatilities
usually is modified.

(ii) By ranking, we exchange the values of the resulting series from
step (i) with that of the original record. The largest value in the
resulting series is replaced by the largest one in the original
series, the second largest value is replaced by the second largest
one, and so on. This step restores the original distribution but now
the power spectrum is changed.

To achieve the convergence to the desired power spectrum and
distribution, we repeat these two steps 30 times. By this way, a
``surrogate'' series is generated. Because of the Wiener-Kinchine
theorem~\cite{Kampen92}, the surrogate record has the same linear
correlations as the original, as well as the distribution. The only
difference is that the original record has the non-linear correlations
(if they exist) but the surrogate does not have any non-linear
features.

In Fig.~\ref{Fig1} we also plot the cumulative distribution for the
surrogate with the same three thresholds as the original. Since the
surrogate records lost the non-linear correlations, they are similar
to each other, we only show results for GE's surrogate. It is seen
that the collapse of the surrogate for different $q$ values is
significantly better than that of the original and the deviation
tendency with the threshold in the original records disappears. This
indicates that the scaling deviations in the original are due to the
non-linear correlations in the volatility. To further test this
hypothesis, we analyze the moments
$\mu_m\equiv\langle(\tau/\langle\tau\rangle)^m\rangle^{1/m}$ in
Sec. (IV) and show similar and consistent deviations from scaling. We
also compare our results to the stretched exponential distribution
(dashed lines). This curve is very close to the empirical results, in
particular for the surrogate records which contain only the linear
correlations. This suggests that PDF of return intervals is well
approximated by a stretched exponential.

\section{The Moments of Scaled Intervals}

The cumulative distribution shows clear systematic trend with $q$,
which is difficult to see from the PDF
directly~\cite{Yamasaki05,Wang06,Wang07,Altmann05,Bunde04,Bunde05}.
To further analyze the systematic tendency in the distribution, we
calculate the moments $\mu_m$ averaged over a stock dataset as a
function of $\langle\tau\rangle$, where a mean interval
$\langle\tau\rangle$ corresponds to a threshold $q$ and therefore
characterizes a return interval series. We study moments for a wide
range of $\langle\tau\rangle$, from $3$ minutes (to avoid the
artificial effects due to discreteness close to $\tau=1$) to thousands
minutes (few trading days or even a week). Assuming a single scaling
function for the PDF $P_q(\tau)$, Eq.~(\ref{pdf.eq}), it follows
\begin{equation}
\mu_m\equiv\langle(\tau/\langle\tau\rangle)^m\rangle^{1/m}=\{\int^\infty_0
(\tau/\langle\tau\rangle)^m\cdot \frac{1}{\langle\tau\rangle}
f(\tau/\langle\tau\rangle) d\tau\}^{1/m}=\{\int^\infty_0 x^m f(x)
dx\}^{1/m},
\label{moment.eq}
\end{equation}
which only depends on $m$ and on the form of the scaling function
$f(x)$ but independent of $\langle\tau\rangle$. Thus, if $\mu_m$
depends on $\langle\tau\rangle$, it suggests deviation from the
assumption of scaling.

\subsection{Moments vs. Mean Interval $\langle\tau\rangle$}

\begin{figure*}
\begin{center}
   \includegraphics[width=0.65\textwidth, angle = -90]{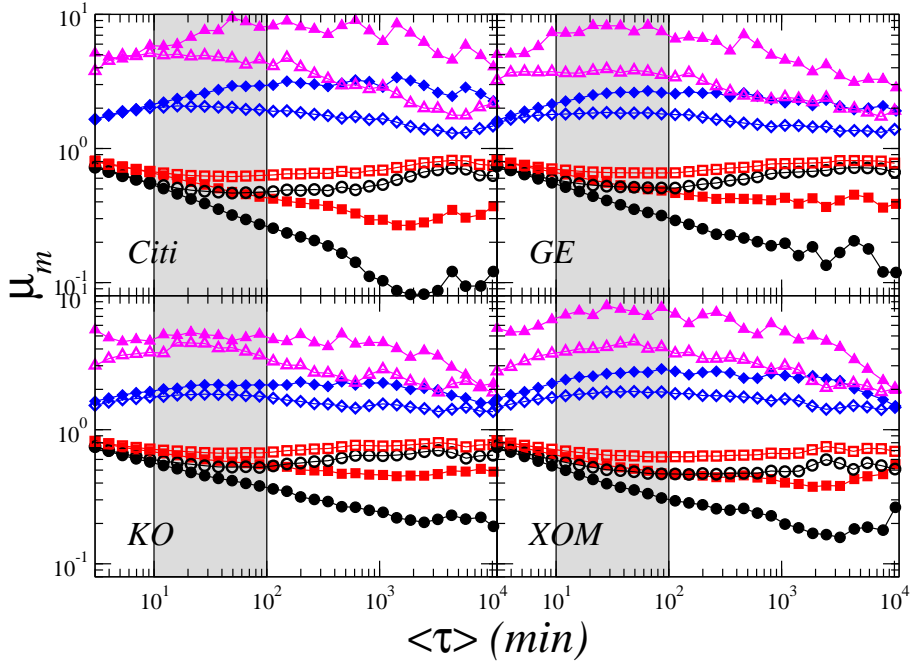}
\end{center}
\caption{(Color online) Moment $\mu_m$ for the scaled intervals of
stock C, GE, KO and XOM. We show results from the original volatility
series (filled symbols) and their surrogate (empty symbols). For each
case, four moments, $m=0.25$ (circles), $0.5$ (squares), $2$
(diamonds) and $4$ (triangles) are demonstrated. Symbols for the
original are clearly away from the horizontal line and their
deviations are much larger than that for surrogate, therefore the
non-linear correlations in the original are related to those
deviations. To avoid effects from the resolution and size limit, we
choose the shadow area, $10<\langle\tau\rangle\leq100$, to study the
multiscaling behavior.}
\label{Fig2}
\end{figure*}

First we examine the relation between the moments $\mu_m$ and the mean
interval $\langle\tau\rangle$. Fig.~\ref{Fig2} shows four
representative moments $m=0.25$, $0.5$, $2$ and $4$ for stock C, GE,
KO and XOM. Ignoring small fluctuations, which is usually due to
limited size data, all moments $\mu_m$ for the original records
deviate significantly from a horizontal line, which is expected for a
perfect scaling of the PDF. They depend on $\langle\tau\rangle$ and
show some systematic tendency. For $m>1$, moments have similar convex
structure, first $\mu_m$ increases with $\langle\tau\rangle$ and then
decreases, where the crossover starts earlier for larger $m$. For
$m<1$, moments also show similar tendency but in the opposite
direction compared to $m>1$. These deviations from scaling in $\mu_m$
are consistent with the deviations seen in the cumulative
distributions shown in Fig.~\ref{Fig1}. Moments of large $m$ ($m>1$)
represent large $\tau/\langle\tau\rangle$ in the PDF and they
initially (for $\langle\tau\rangle\leq100$) increase with
$\langle\tau\rangle$, while moments of small $m$ ($m<1$) represent
small $\tau/\langle\tau\rangle$ and they initially (for
$\langle\tau\rangle\leq100$) decrease with $\langle\tau\rangle$.

To further test if the systematic deviations are not due to finite
size effects and discreteness, we also examine moments for the
surrogate records which are more flat for most range, as shown in
Fig.~\ref{Fig2}. For the same order $m$, the moment of the surrogate
obviously differs from that of the original, especially in the medium
range of $\langle\tau\rangle$ ($10<\langle\tau\rangle\leq100$). This
discrepancy suggests that the non-linear correlations exist in the
original volatility and accounts for the scaling
deviations. Nevertheless, all moments of surrogate show small
curvature from a perfect straight line at both short and long
$\langle\tau\rangle$, which are much weaker compared to the original
records. The weak curvature suggests that some additional effects, not
related to the non-linear correlations, affect the moments. For small
$\langle\tau\rangle$, the resolution discrete limit seems to have some
influence on the moments. We will discuss this effect in section
C. For large $\langle\tau\rangle$, the moments are gradually
approaching the horizontal line and are more fluctuating, the effect
seems to be related to limited size of the record. This effect will be
discussed in section B.

\subsection{Multiscaling}

For the original volatility records, the systematic tendency in the
distribution of $\tau$ and the moments implies that the return
intervals may have multiscaling features. To avoid as much as possible
the effect of discreteness and finite size, we calculate the moments
only for some medium range of $\langle\tau\rangle$ where the effects
are small. Since there is no non-linear correlations in the surrogate
records, the curvature in their moments is only due to the additional
effects, we use the surrogate curve as our reference. For small
$\langle\tau\rangle$, the increasing (decreasing) range for $m>1$
($m<1$) almost ends at $\langle\tau\rangle=10$ minutes. For large
$\langle\tau\rangle$, the curves start to decrease (increase) from
different positions, but at $\langle\tau\rangle=100$, all curves do
not or just start to decrease (increase). Thus we choose to study
$\mu_m$ in the region, $10<\langle\tau\rangle\leq100$, represented by
the shadow areas in Fig.~\ref{Fig2}. In this range, we find a clear
trend for the original records while the surrogate is almost
horizontal. To quantify the tendency, we fit the moments with a
power-law,
\begin{equation}
\mu_m \sim \langle\tau\rangle ^ \alpha.
\label{alpha.eq}
\end{equation}
If the distribution of $\tau/\langle\tau\rangle$ follows a scaling
law, the exponent $\alpha$ should be some value very close to $0$. If
$\alpha$ is significantly different from $0$, it suggests
multiscaling.

\begin{figure*}
\begin{center}
   \includegraphics[width=0.65\textwidth, angle = -90]{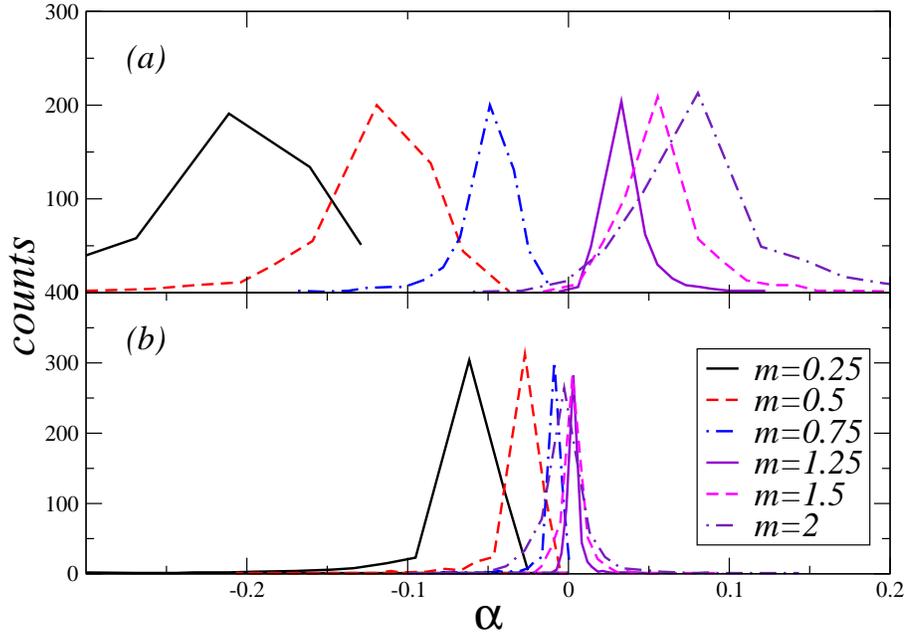}
\end{center}
\caption{(Color online) Distribution of multiscaling exponent $\alpha$
for S\&P 500 constituents. The exponent $\alpha$ is obtained from the
power-law fitting for moments in the medium range
$10<\langle\tau\rangle\leq100$. (a) Histogram of $\alpha$ for the
original volatility and (b) for surrogate. The distributions have a
systematic shift with $m$ in (a) while all of them almost collapse in
(b). This suggests the multiscaling behavior in the return intervals
in the original records. The significant discrepancy between the
original and the surrogate records manifests that non-linear
correlations form the original volatility accounts for the
multiscaling behavior.}
\label{Fig3}

\end{figure*}

To examine the multiscaling behavior for the whole market, we
calculate $\alpha$ for all 500 stocks of S\&P 500 constituents and
plot the histogram for $m=0.25$ to $2$. Fig.~\ref{Fig3} shows that
each histogram has a narrow distribution, which suggests that $\alpha$
are similar for the 500 stocks. For the original records, almost all
$\alpha$ significantly differ from $0$, thus the moments clearly
depend on the mean interval. Moreover, the mean value of $\alpha$
shifts with order $m$ from $\langle\alpha\rangle\simeq-0.2$ for
$m=0.25$ to $\langle\alpha\rangle\simeq0.1$ for $m=2$ which means the
dependence varies with the order $m$. This behavior suggests
multiscaling in the return intervals distribution. Indeed, histograms
for the surrogate records are more centered around values close to
$\alpha=0$. The uniscaling behavior for the surrogate suggests that
the non-linear correlations in the volatility are responsible for the
multiscaling behavior in the original.

\begin{figure*}
\begin{center}
   \includegraphics[width=0.65\textwidth, angle = -90]{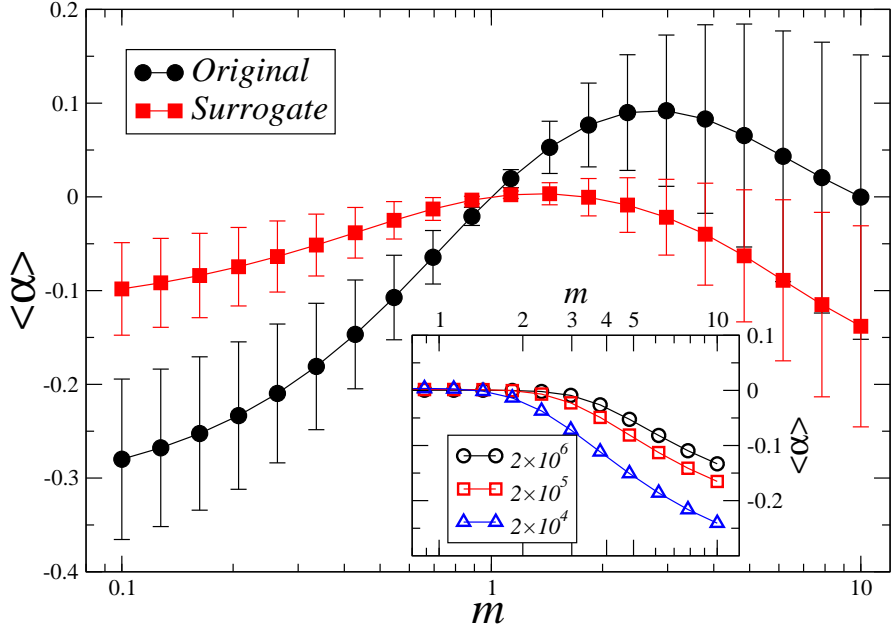}
\end{center}
\caption{(Color online) Dependence of average multiscaling exponent
$\langle\alpha\rangle$ on order $m$. The average
$\langle\alpha\rangle$ was taken over the 500 $\alpha$ of the S\&P 500
constituents. Results for the original (circles) and surrogate records
(squares) are displayed. For large $m$, the two curves have the
similar tendency which is attributed to the finite size effects. For
small $m$, the two curves are significantly different which supports
the multiscaling in the return intervals, due to the non-linear
correlations in the original volatility. The inset demonstrates
$\langle\alpha\rangle$ averaged over 500 stretched exponential
distributed i.i.d. simulations with $\gamma=0.3$. Three sizes,
$2\times 10^6$, $2\times 10^5$ and $2\times 10^4$ are displayed, which
clearly shows the finite size effect.}
\label{Fig4}
\end{figure*}

To remove fluctuations and show the tendency clearer, we plot the
dependence of $\langle\alpha\rangle$ on $m$, where
$\langle\alpha\rangle$ is the average $\alpha$ over all 500 stocks. In
Fig.~\ref{Fig4} we show this relation for a wide range of $m$,
$0.1\leq m\leq 10$, and the plot shows two different behaviors. For
small $m$ (roughly $m\leq2$), $\langle\alpha\rangle$ for the original
records clearly deviates from 0 and demonstrates the multiscaling
behavior, while $\langle\alpha\rangle$ for the surrogate is closer to
$0$. For large $m$ ($m>2$), the two curves have similar decreasing
trend. Since large $\tau$ dominates high order moments, this
similarity may be due to finite size effects. To test the finite size
effects we simulate surrogate return intervals by assuming a stretched
exponential distribution i.i.d. process with 3 sizes (number of all
$\tau$ in the series), $2\times 10^6$, $2\times 10^5$ (the size of the
empirical dataset) and $2\times 10^4$. Without loss of generality, we
choose $\gamma=0.3$, which is the correlation exponent for GE of
$q=2$. To be consistent with the 500 stocks, we perform 500
realizations and plot their average exponent
$\langle\alpha\rangle$. As shown in the inset of Fig.~\ref{Fig4}, all
$\langle\alpha\rangle$ show a similar decreasing trend as that of the
empirical curves. However, it is seen that the trend starts earlier
for smaller size, and thus the size limit has a strong influence on
high order moments. Fig.~\ref{Fig4} also shows the error bars for the
two records, which is the standard deviation of 500 $\alpha$
values. Note that the error bars for the volatility records do not
overlap those of their corresponding surrogate, indicating the
significance of our results.

\subsection{Discreteness Effect}

\begin{figure*}
\begin{center}
   \includegraphics[width=0.65\textwidth, angle = -90]{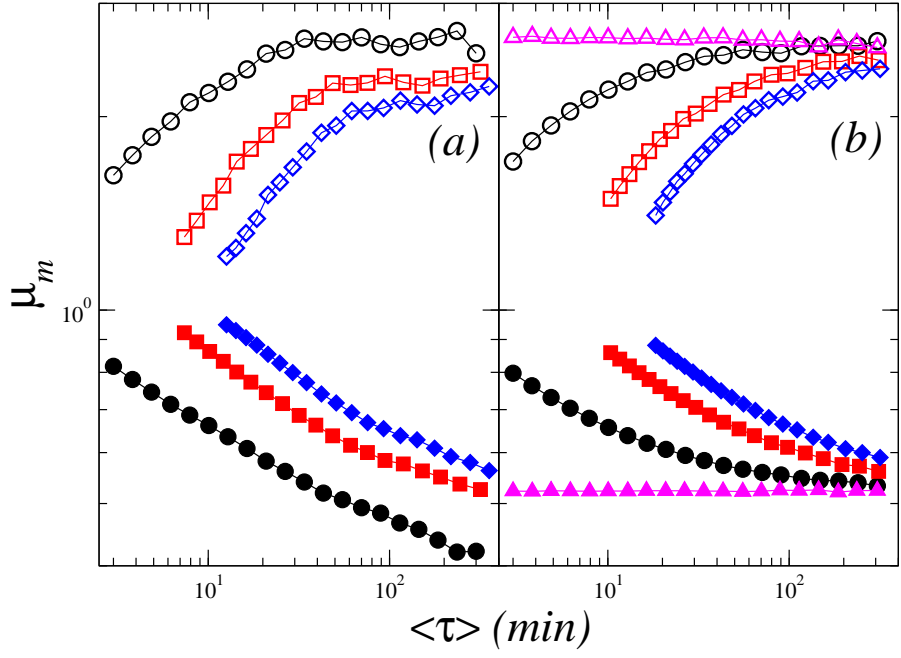}
\end{center}
\caption{(Color online) Discreteness effect in the moments. (a)
Moments for stock GE with three resolutions, 1 (circles), 5 (squares)
and 10 minutes (triangles). Filled symbols are for $m=0.5$ while empty
symbols are for $m=2$. (b) Moments of artificial records averaged over
100 simulations. For each trial, we simulate the return intervals
which follows a stretched exponential distribution with $\gamma=0.3$
and the length of 200 thousands points. Symbols are similar to that in
(a). Three resolutions, 1, 5 and 10 time units are displayed. To show
the disappearing of the discreteness, we also plot simulation results
for continuous return intervals (triangles), which exhibits a constant
moment, independent of $\langle\tau\rangle$.}
\label{Fig5}
\end{figure*}

For small $\langle\tau\rangle$ ($\langle\tau\rangle\le10$), the
behavior of $\mu_m$ as a function of $\langle\tau\rangle$ was
attributed to the discreteness. Here we examine this effect. Due to
the limits in recording, we can not have a continuous but discrete
record. In our study the volatility is recorded in 1 minute. The
relative errors in moments will be considerable large for small
$\langle\tau\rangle$ close to $1$ minute. By starting from
$\langle\tau\rangle=3$, we only partially avoid the discreteness in
the moments. To test the discreteness effects, since we can not
increase the resolution, we reduce it and compare the moments $\mu_m$
with 3 resolutions, 1 minute, 5 minutes and 10
minutes. Fig.~\ref{Fig5}(a) shows this comparison for GE for $m=0.5$
and $m=2$. The three resolutions have a similar trend, showing that
the curves become flatter for the higher resolution. For other stocks,
we find similar behavior. This systematic tendency suggests that the
recording limit (1 minute) strongly affects the moments at short
$\langle\tau\rangle$. To reduce it, we should raise the recording
precision or study the moments of larger $\langle\tau\rangle$.

To further test this result, we simulate artificial return intervals
with an i.i.d. process from stretched exponential distribution with
$\gamma=0.3$, same as the empirical $\gamma$ for GE of $q=2$. We
examine moments of $m=0.5$ and $m=2$ with the same three resolutions
(1, 5 and 10 time units) as in the empirical test done above. The
simulated size is 200 thousands points for each trial and we use the
average over 100 trials for each resolution. Fig.~\ref{Fig5}(b) shows
curves similar to that of empirical (Fig.~\ref{Fig5}(a)). For the
higher resolution, the curve is closer to the horizontal line and
finally may reach the line when we raise the resolution high
enough. To show this, we also simulate continuous return intervals and
find constant moments, as expected (Fig.~\ref{Fig5}(b)). Therefore,
the discreteness effects can be overcome if the resolution is improved
enough. Note that for the empirical data, we expect the moment not to
be constant for small $\langle\tau\rangle$ even if we have a much
better resolution, since the return intervals has the multiscaling
behavior, as shown for larger $\langle\tau\rangle$ in the range
$10<\langle\tau\rangle\leq100$ which is not affected by discreteness.

\subsection{Moments vs. Order $m$}

\begin{figure*}
\begin{center}
   \includegraphics[width=0.65\textwidth, angle = -90]{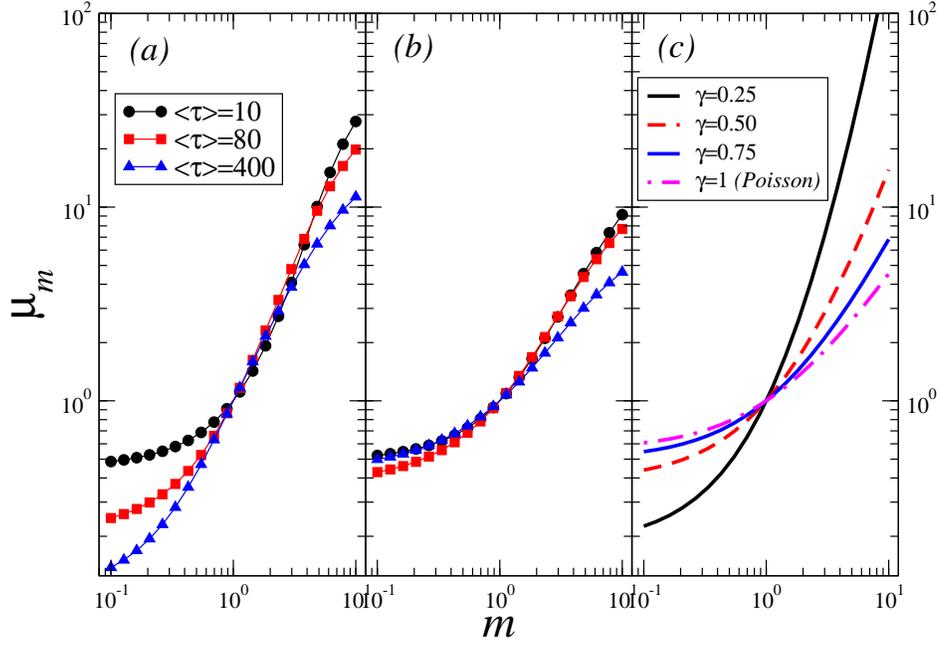}
\end{center}
\caption{(Color online) Dependence of the moment $\mu_m$ on the order
$m$. (a) for the return intervals of the original volatility for stock
GE. (b) for its surrogate. Three mean interval $\langle\tau\rangle=10$
(circles), $80$ (squares) and $400$ minutes (triangles) are
demonstrated in (a) and (b). (c) Analytical moments from stretched
exponential distributions, $\gamma=0.25$, $0.50$, $0.75$ and $1$
(Poisson distribution) taken from Eq.~(\ref{moment1.eq}). For large
$m$, both (a) and (b) show discrepancies which is related to the
finite size effect. For small $m$, the difference between (a) and (b)
is due to the non-linear correlations in the original
volatility. Compared to the analytical curves in (c), the return
intervals for the original records shows multiscaling behavior.}
\label{Fig6}
\end{figure*}

The moments $\mu_m$ have systematic dependence on $m$, as seen in
Figs.~\ref{Fig2} and ~\ref{Fig3} where the moments are plotted as the
function of $\langle\tau\rangle$. It is of interest to explore the
relation between the moments and $m$ directly. For a fixed
$\langle\tau\rangle$, representing a given threshold $q$ one can
study, the return intervals and their moments of various orders which
exhibit information on different scales of $\tau$. Moments of large
$m$ represent large $\tau$ and vice versa. If
$\tau/\langle\tau\rangle$ follows a single distribution without
corrections due to effects such as discreteness and finite size,
curves of $\mu_m$ vs. $m$ for different $\langle\tau\rangle$ should
collapse to a single one, which only depends on the scaling function
$f(x)$ from Eq. (\ref{moment.eq}). In Fig.~\ref{Fig6} we plot $\mu_m$
vs. $m$ for both the original and surrogate records. We plot $\mu_m$
for $m$ between $0.1$ and $10$ for three $\langle\tau\rangle$ values:
10, 80 and 400 minutes. For the original (Fig.~\ref{Fig6}(a)), there
is substantial deviations from a single curve. This supports our
suggestion that the return intervals has multiscaling
behavior. Moments for the surrogate (Fig.~\ref{Fig6}(b)) converge to a
single curve for $m\le2$ but become diverse for high orders, which
agrees with the strong influence of the finite size effects. As a
reference, we also plot the analytical moments (Fig.~\ref{Fig6}(c))
from the stretched exponential distribution. Substituting
Eq.~(\ref{scaling.eq}) into Eq.~(\ref{moment.eq}), we obtain
\begin{equation}
\mu_m=\frac{1}{a}\{\frac{\Gamma((m+1)/\gamma)}{\Gamma(1/\gamma)}\}^{1/m}.
\label{moment1.eq}
\end{equation}
 Fig.~\ref{Fig6}(c) shows analytical curves for various correlation
exponent $\gamma$.

\section{Discussion}

We study the scaling properties of the distribution of the volatility
return intervals for all S\&P 500 constituents. We find small but
systematic deviations from scaling assumption with the threshold $q$
in the cumulative distribution. Compared to the good collapse for the
surrogate records where non-linearities are removed, this suggests
that the origin of this trend is due to non-linear correlations in the
original volatility. Moreover, we find similar systematic deviations
for the moments $\mu_m$, which are also attributed to the non-linear
correlations in the volatility. We distinguish these deviations from
the deviations due to the discreteness for small $\langle\tau\rangle$
and finite size effect for large $\langle\tau\rangle$. Further, we
explore the dependence of the moment $\mu_m$ on its order $m$. When
compare to surrogate records and to analytical curves, the results
support the multiscaling hypothesis of the return intervals. Thus, the
scaling assumption in the return interval distributions although it is
a good approximation can not be exact. Also, the stretched exponential
form of the scaling function can only be an approximation. Recently
Eisler et al. \cite{Eisler06A, Eisler06B} exhibits that the
distribution of intertrade times has similar multiscaling behavior and
the market activity depends on the company capitalization. It would be
interesting to connect the intertrade times with the return intervals
and test size dependence in the return intervals.

\section*{Acknowledgments}

We thank Y. Ashkenazy for his kind help in the simulations,
R. Mantegna, J. Kert\'{e}sz and Z. Eisler for fruitful discussions,
and the NSF and Merck Foundation for financial support.

\newpage


\end{document}